\documentclass[superscriptaddress,cite,prl,twocolumn]{revtex4-1}
\usepackage{epsfig,pslatex,latexsym,times,amssymb,amsmath,graphicx,bm,revsymb}
\pdfoutput=1
\begin{document}
\author{M. Prada}
\email{mprada@physnet.uni-hamburg.de}
\affiliation{I. Institut f\"ur Theoretische Physik, Universit\"at Hamburg, Jungiusstr. 9, 20355 Hamburg, Germany}

\date{\today}
\title{ 
The geometric phase of Z$_n$- and  T-symmetric nanomagnets  as a classification toolkit 
} 
\newcommand*{\ud}{\mathrm{d}}
\newcommand*{\ue}{\mathrm{e}}

\begin{abstract}
{
We derive the general form of the non-trivial geometric phase resulting from  the unique combination of point group and time 
reversal symmetries. 
This phase arises {\it e.g.} when a magnetic adatom is adsorbed on a non-magnetic C$_n$ crystal surface, 
where $n$ denotes the fold of the principal axis. 
The energetic ordering and the relevant quantum numbers of the eigenstates are entirely determined by  this quantity. 
Moreover, this phase allows to conveniently predict the protection mechanism of any prepared state, shedding 
light onto a large number of experiments and allowing a  classification scheme. 
Owing to its robustness this geometric phase  also has  great relevance for a large number of applications in 
 quantum computing, where topologically protected states bearing long relaxation times are highly desired. 
}
\end{abstract}


\maketitle

\section{\bf Introduction} 
In the early 80's, Berry discovered an intriguing, non-integrable phase depending only on the geometry of the parametric space \cite{berry}. 
This phase, which had been overlooked for decades, provided a deep insight on the geometric structure of quantum mechanics, resulting in 
various observable effects. 
The concept of the Berry phase is a central unifying concept in quantum mechanics, 
shedding light onto a broad range of phenomena such as
the Aharonov-Bohm effect, the quantum and the anomalous Hall effect, etc. 
Moreover, geometric or Berry phases nowadays represent the most robust resource for storing and processing quantum information \cite{guido}. 
 
In this work, we focus on the fundamental aspects of the geometric phase arising from the combination of $n$-fold (Z$_n$) point group
and time reversal (TR) symmetries. 
This concept appears in a clean and illustrative form in spin adsorbates (SA) on non-magnetic C$_n$ symmetric crystal structures
\cite{qc1,nachoT,nacho4,ssm,ssm3,loth,chistou,brune,nacho3,delgado,donati2,ternes,nacho1,rossK,nico2,ternes2}. 
Owing to the spin orbit coupling, these systems possess $n$-fold spin rotations while being time-reversal symmetric.
The SA's large magnetic moment 
adapts  to the anisotropic crystal field , resulting in non-trivial spin dynamics  and 
bearing enhanced lifetimes of the two degenerate ground states in the SA \cite{adatoms1,adatoms2,adatoms3,NatK}. 
However, a general classification scheme of this novel symmetry combination has not been so far reported, nor a comprehensive 
study of the mechanisms responsible for this spin protection from a general perspective. 

Here we derive an exotic geometric phase that captures all the possible symmetry combinations of 
a $n$-fold rotational magnet. 
With only this quantity,  a general classification scheme based purely on symmetry arguments follows. 
In addition, we obtain two quantum numbers resulting from the intriguing combination of the symmetries. 
Finally, we are able to predict the allowed transitions induced by a general exchange interaction, finding 
topologically protected states for some particular combinations of spin and $n$. 
This  phase permits  thus to deal with two related problems at once, 
both relevant to the use single quantized spins to store classical information: 
whether or not the ground state is doubly degenerate, and 
whether or not direct exchange-mediated spin transitions may occur between the 
two ground  states. 
Hence, this phase may suggest novel designs of fault-tolerant, SA based quantum logic gates.

\section{\bf Methods} 
A free SA is characterized 
by a set of $(2J+1)$ fully degenerate eigenstates, $\{|J,M\rangle\}$, 
where $J$ denotes the {`spin', {\it e.g.} the effective spin $\tilde S$ for transition metal 
ions, or the total angular momentum $L+S$ for rare earth ions}, and $M$ denotes its $z$-component. 
{The symmetry is lowered when the SA is placed on an axially symmetric site, and it may be described by a generic axial term 
$\hat H_{\rm u} = D\hat{J}_z^2$, resulting in a set of $J$ ($J+1/2$) pairs of states for $J$ integer (half-integer). 
For transition metal ions, $\hat H_{\rm u}$ represents the axial zero field splitting  term of an effective spin Hamiltonian, 
whereas for rare earths, $\hat H_{\rm u}$ is a physical crystal field Hamiltonian \cite{rudowicz5}.} 
{For rhombic symmetry sites, a generic rhombic term $\hat H_n$  must be added,}
which mixes the magnetic quantum numbers and hence $M$ ceases to be a good quantum number.
$\hat H_n$ is a combination of $n$-powers of the ladder operators, \cite{book,NatK}, 
$\hat J^\pm = \hat J_x \pm i\hat J_y$. 
To lowest order, the generic Hamiltonian for the SA reads:
\begin{equation} 
\hat H = \hat H_{\rm u} + \hat H_n  = D_z(\hat J_z)^2 + E_n\{[(\hat{J}_+)^n + (\hat{J}_-)^n],\hat{J}_z^\alpha\}, 
\label{eqHJ}
\end{equation}
with $\{,\}$ denoting anticommutator.  $\alpha$ = 0 (1) for even (odd) $n$ to ensure TR invariance of $H$ \cite{schuh}.
{The generic Hamiltonian in Eq. (1) can equivalently be expressed in terms of extended Steven’s operators \cite{stevens} 
$\hat O^q_k(\hat X)$, with $\hat X = \hat J (\hat S)$ in crystal field (zero-field splitting) Hamiltonian  \cite{wybourne}, 
where the meaning of `spin' should not be confused \cite{misra,rudowiczC,rudowicz5}.

Owing to the discrete rotational and time-reversal symmetries associated with the point group of the crystal C$_n$ 
and the absence of an external magnetic field, respectively, we obtain:
\begin{eqnarray}
\label{eqHR0}
[\hat H,\hat R_{2\pi /n}^z] &=& 0, \\ 
\label{eqHT0}
[\hat H,\mathcal{ \hat T}] &=& 0, 
\end{eqnarray}
where we have defined the rotations of the C$_n$ point group 
$\hat R_{2\pi /n}^z$, and the TR operator $\mathcal{ \hat T}$. 
The irreducible representation of the $n$-dimensional C$_n$ group is given 
by $\{1, e^{i2\pi /n}, \dots e^{i2\pi(n-1) /n}\}$, whereas the irreducible representation generated by 
TR has two elements, $\{1,-1\}$. 
Hence, it appears natural to label the eigenstates
according to the phase acquired under a symmetry spin rotation, 
$2\pi m/n$ and the one related to the time-reversal operator, $\varsigma(m,n)$.  
We are thus introducing two new quantum numbers: $m$, arising from the 
discrete rotations of the $C_n$ group, and $\varsigma(m,n)$, describing the TR transformations.
Considering first the rotations, we have
\begin{equation}
\hat R_{2\pi /n} |\psi_{m}^{\varsigma}\rangle = e^{i2\pi m/n}|\psi_m^{\varsigma}\rangle.
\label{eqRop}
\end{equation}
On what follows we choose $m$, an arbitrary representative of the rotational subgroup, 
to be the element with largest absolute value of $z$-component of the spin. 
We note that this rotationally invariant representation would correspond to the 
coloring code of the states commonly being used \cite{NatK, huebner,karlewski}, 
and we want to point out its relation to the discrete $2\pi/n$-rotations. 
This connection does not appear clear in the present literature.
Finally, $|\psi_{m}^{\varsigma}\rangle$ is a linear combination of $|J, M\rangle$ states \cite{book}, 
\begin{equation}
|\psi_{\pm m}^{\varsigma}\rangle = \sum_{k:|m-n k|\leq J} c_k^{n} |J,\pm( m -n k)\rangle. 
\label{eqpsim}
\end{equation}
$c_k^m$ are some coefficients to be determined by the environmental field \cite{misra}, and the sum is evaluated 
over a set of  $k$ integer with the condition $|m-n k|\leq J$.
We can immediately see that $|\psi_{\pm m}^{\varsigma}\rangle $ satisfies (\ref{eqRop}),
\begin{eqnarray}
\hat R_{2\pi /n} |\psi_{\pm m}^{\varsigma}\rangle &=& \sum_{k:|m-n k|\leq J} c_k^{n} \hat R_{2\pi /n} |J,\pm( m -n k)\rangle =
\nonumber \\ &=&
\sum_{k:|m-n k|\leq J} c_k^{n} e^{i(m-nk)2\pi/n} |J,\pm( m -n k)\rangle =
\nonumber \\ &=&
 e^{i2\pi m/n}|\psi_m^{\varsigma}\rangle. \nonumber
\end{eqnarray}
For the $c_k^m$ coefficients, we may expect the largest contribution to the ground state to be given by $k = 0$ at $m = \pm J$, in the 
typical case where  $|D_z| \gg |E_n|$ and $D_z<0$. 
$\varsigma$ labels the TR phase that connects  two states in the time-reversed pair. 

We now focus on the action of $\mathcal{ \hat T}$ in (\ref{eqpsim}).  Following the procedure carried out in \cite{NatK} (supplemental material), we obtain:  
\begin{align}
\mathcal{ \hat T}|\psi_{ m}^{\varsigma}\rangle &
= e^{i\pi m} \sum_{k:|m-n k|\leq J} (c_k^m)^* e^{-i\pi n k} |J,-( m -n k)\rangle.
\label{eqtr1}
\end{align}
Applying the operator twice, $\mathcal{ \hat T}^2$, we 
obtain the usual eigenvalues $\pm 1$ (positive for integer $m$ and negative, for $m$ half-integer. 
However, noting the axial nature of spin,  we have that  $\mathcal{ \hat T} \hat R_{2\pi/n}^z = \hat R_{-2\pi/n}^z \mathcal{ \hat T}$, which involves
that  $\mathcal{ \hat T}$ and $ \mathcal{ \hat T^-} \equiv \hat R_{-2\pi/n}^z\mathcal{ \hat T} \hat R_{-2\pi/n}^z$ are two alternative `paths' of the 
time reversal operator (see Fig. \ref{cartoon}). 
Likewise, we may define its counterpart: 
$\mathcal{ \hat T^+} \equiv \hat R^z_{2\pi /n} \mathcal{ \hat T} \hat R^z_{2\pi /n} $, involving anti-clockwise rotations. 
Hence, the action of TR and this new `rotated time reversal' (RTR) operators should yield the same state up to a global phase. 
We may thus define this phase as our quantum number $\varsigma(m)$ related to the RTR symmetry, in analogy with the TR operation:
\begin{eqnarray}
 \mathcal{ \hat T^-} |\psi_{m}^{\varsigma(m)}\rangle &=& e^{ i\varsigma(m)}|\psi_{-m}^{\varsigma(-m)}\rangle; \quad 
\nonumber \\ 
(\mathcal{ \hat T^-})^2 |\psi_{m}^{\varsigma}\rangle  &=&  \mathcal{ \hat T^-} e^{i\varsigma(m)}|\psi_{-m}^{\varsigma(-m)}\rangle = 
e^{-i[\varsigma(m) -\varsigma(-m)]}|\psi_{m}^{\varsigma(m)}\rangle. 
\label{eqTRop}
\end{eqnarray}
An important point to note is that the two  quantum numbers related to the symmetries arising as a consequence of 
(\ref{eqHR0}) and (\ref{eqHT0}), are mutually dependent in the conventional representation, where 
the action of spin rotations is given by (\ref{eqRop}). This action depends on the representative $m$ and the fold of the axis $n$, 
which implies that the action of the TR {\it depends as well on both parameters}, $m$ and $n$. 
That is, once  $n$ is fixed, $\varsigma$ is a function of $m$. 
Without loosing generality, we may choose $\varsigma(m)= -\varsigma(-m)$, bearing:  
\begin{equation}
(\mathcal{ \hat T^-})^2 |\psi_{m}^{\varsigma}\rangle =
e^{-2i\varsigma(m)}|\psi_{m}^{\varsigma(m)}\rangle.
\label{eqT2}
\end{equation}
Bearing in mind the anti-unitarity of $\mathcal{ \hat T^-} $ and using (\ref{eqRop}) and (\ref{eqtr1}), we encounter: 
\begin{widetext}
\begin{eqnarray*}
(\mathcal{ \hat T^-})^2  |\psi_{m}^{\varsigma}\rangle  &=& 
\hat R^z_{-2\pi /n} \mathcal{ \hat T} \hat R^z_{-2\pi /n} \hat R^z_{-2\pi /n} \mathcal{ \hat T} \hat R^z_{-2\pi /n} |\psi_{m}^{\varsigma}\rangle = 
\hat R^z_{-2\pi /n} \mathcal{ \hat T} \hat R^z_{-4\pi /n} \mathcal{ \hat T}e^{-i2\pi m/n} |\psi_{m}^{\varsigma}\rangle = 
\nonumber \\ &=& 
\hat R^z_{-2\pi /n} \mathcal{ \hat T} \hat R^z_{-4\pi /n} e^{i2\pi m/n} e^{i\pi m} \sum_{k:|m-n k|\leq J} (c_k^m)^* e^{-i\pi n k} |J,-( m -n k)\rangle =
\nonumber \\ &=& 
\hat R^z_{-2\pi /n} \mathcal{ \hat T} e^{i6\pi m/n}e^{i\pi m} \sum_{k:|m-n k|\leq J} (c_k^m)^* e^{-i\pi n k} |J,-( m -n k)\rangle = 
\nonumber \\ &=& 
 e^{-i8\pi m/n}e^{-2i\pi m}  \sum_{k:|m-n k|\leq J} (c_k^m) |J,( m -n k)\rangle = e^{-2i\pi m(1+4/n) }|\psi_{m}^{\varsigma}\rangle, 
\label{eqtminus0}
\end{eqnarray*}
\end{widetext}
Comparing (\ref{eqtminus0}) with the definition in (\ref{eqT2}), we obtain the global phase $\varsigma$: 
\begin{equation}
\varsigma(m,n) 
= \pi m \left(1 + \frac{4}{n}\right). 
\label{eqBerryPhase}
\end{equation}
We note that a similar result can be obtained by the action of an `anti-clockwise' rotation,  
$ \mathcal{ \hat T^+}\equiv \hat R^z_{2\pi /n} \mathcal{ \hat T} \hat R^z_{2\pi /n} $. 
It is straight forward to perform a similar derivation as in (\ref{eqTRop}-\ref{eqtminus0}), resulting in an alternative definition of the 
global phase, 
\begin{eqnarray}
(\mathcal{ \hat T^+})^2  |\psi_{m}^{\varsigma^\prime(m)}\rangle &=& 
\mathcal{ \hat T^+}e^{i\varsigma^\prime(m) }|\psi_{m}^{\varsigma^\prime(m)}\rangle =
e^{-2i\varsigma^\prime(m) }|\psi_{m}^{\varsigma^\prime(m)}\rangle=
\nonumber \\ &=&
e^{2i\pi m(1+4/n) }|\psi_{m}^{\varsigma^\prime(m)}\rangle, 
\label{eqtplus}
\end{eqnarray}
with \[
\varsigma^\prime (m,n)
= \pi m \left(1 - \frac{4}{n}\right), \]
which would result in a similar classification scheme. 
In general, we may write: 
\begin{equation}
\mathcal{ \hat T^+} |\psi_{m}^{\varsigma(m)}\rangle = 
e^{i\pi m(1- 4/n)}
|\psi_{-m}^{\varsigma(-m)}\rangle = 
e^{i\varsigma(m)}
|\psi_{-m}^{\varsigma(-m)}\rangle. 
\label{eqTpp}
\end{equation}
In this sense, we may describe a RTR operation by either expression of $\varsigma$ given in (\ref{eqBerryPhase}) or by that of (\ref{eqtplus}), 
since for each positive representative $m$, its counterpart $-m$ can be chosen as representative. 
Without loosing generality, we have chosen the latter in this work. 
Note that  in the last step of (\ref{eqtplus}) we have used $e^{2i\pi m }= e^{-2i\pi m } \ \forall \ m$. 
With this choice, the action of subsequent $( \mathcal{ \hat T^-})$ and $( \mathcal{ \hat T^+})$ yields the usual Wigner's phase, $2i\pi m$: 
$( \mathcal{ \hat T^+})( \mathcal{ \hat T^-}) |\psi_{m}^{\varsigma} \rangle 
= \hat R^z_{2\pi /n} (\mathcal{ \hat T})^2 \hat R^z_{-2\pi /n} |\psi_{m}^{\varsigma}\rangle = e^{2i\pi m} |\psi_{m}^{\varsigma}\rangle$, 
as required. 
As we will see next, this corresponds to performing a RTR operation and its inverse.  

The action of $\mathcal{ \hat T^+}$ and  $\mathcal{ \hat T^-}$ can be represented by a cyclic group 
of dimension $d = 2|1-4/n|^{-1}$ ($d$=1, for $n=4$) (see Table \ref{table1}), and elements $\{e^{i2\pi i/d}\}$, $i=0,1,\dots d-1$. 
This implies that if $d>n$, the subspace spanned by the representatives $m$ (of dimension $n$) has to be 
enlarged to represent each of the $d$ elements of the subspace generated by the RTR operation.   

\section{ \bf Results and Discussion} 
Eq. (\ref{eqBerryPhase}) is the central result of this work, and can be identified with a geometric  phase
that arises due to the non-trivial TR and rotational symmetry combination. 
\begin{figure}[!htb]
\includegraphics[width=0.35\textwidth]{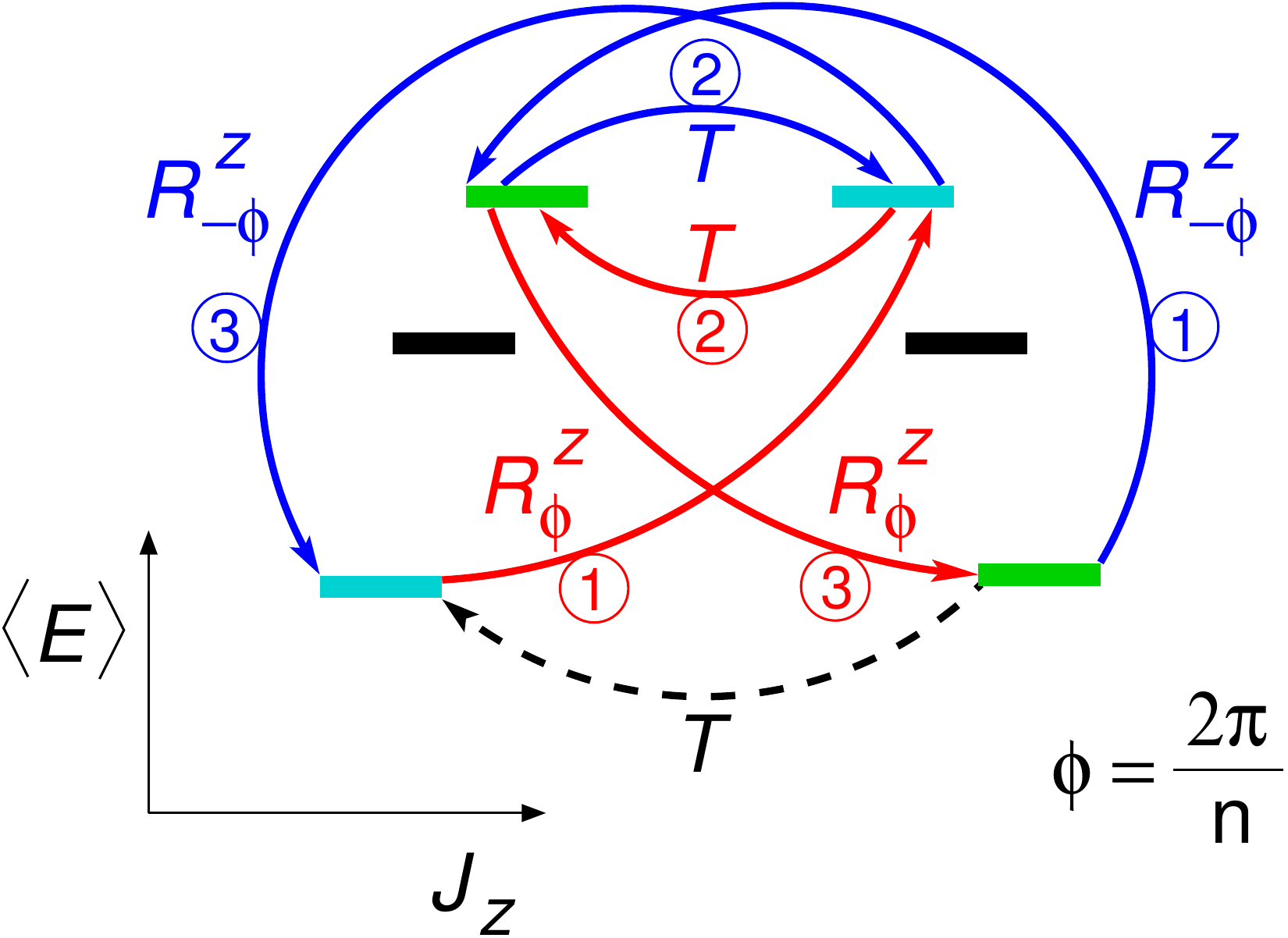}
\caption{Cartoon illustrating the two paths that connect time-reversed pairs for different $J_z$ states 
for $n=3$: 
Direct time-reversal  (black broken arrow) and  rotated time reversal $\mathcal{ \hat T^+}$ 
and $\mathcal{ \hat T^-}$ (red and blue arrows, respectively). 
Note that $J_z$ is {\it {not}} a good quantum number, {\it i.e.}, the symbols do not represent eigenstates, 
although the colors label the rotational class and hence can be connected by symmetry rotations.  
The vertical axis correspond to the mean value of the energy (not to scale). 
}
\label{cartoon}
\end{figure}
A graphical interpretation is given in Fig. \ref{cartoon}, 
where the expectation value of the energy is plotted as a function of $J_z$. Note that 
$J_z$ is not a good quantum number, as the Hamilonian (1) mixes the different $J_z$. 
 $\varsigma(m)$ is the phase acquired 
under a RTR operation for the  states at the bottom of the zone ($n=3$, in this example). 
The red (blue) path is related to $\mathcal{ \hat T^+}$ ($\mathcal{ \hat T^-}$), and yields a global phase of $\pi m(1+4/n)$ ($\pi m(1-4/n)$). 
This phase is topological in the sense that it can only be defined within the cyclic group C$_n$ 
and the related closed paths in the Bloch sphere \cite{loss,delft}. 

To illustrate this theory with an example, we consider the case studied by Miyamachi {\it et al.} \cite{NatK}, 
where they consider  Ho impurities ($J = 8$) on a 3-fold crystal ($n =3$). 
Three rotationally invariant subgroups arise,  
one mixing $M$ = -8, -5, -2, 1, 4, and 7,  
(we may choose as our representative  $m=-8$, usually marked `blue' in literature), 
with rotational phase -8i$\pi$/3 $\equiv$ i$\pi$/3, 
another mixing $M$ = -7, -4, -1, 2, 5, 8  ($m=8$, red), 
with rotational phase 2i$\pi$/3,  
and the last one with $M$ = -6, -3, 0, 3, 6 ($m=6$, green), with rotational phase 0.   
These corresponds to the  irreducible representation  $\left\{ e^{i4\pi/3}, e^{i2\pi/3}, 1\right\}$, respectively, in terms 
of the phase each member of the subgroup acquires under the discrete rotations. 
An equivalent representation, using the representatives $m$, would be  $\{-8,8,6\}$. 
Interestingly, whereas the TR eigenvalue of the isolated adatom is real (recall that $J$ is integer in this specific example), 
the TRT eigenvalue result complex, with dimension $d = 2|1-4/n|^{-1} =6$. 
We then must take $d=2n$ representatives to span the whole subgroups generated by RTR. 
It may be seen by inspection, that the RTR can be generated now by the 6 representatives  $\{-8,8,-7,7,6,3\}$, or equivalently, 
$\{0, 1, -1, 2, -2, 3\}$. These span the 
discrete phases $\left\{e^0, e^{-i\pi/3}, e^{i\pi/3},e^{-2i\pi/3}, e^{2i\pi/3},e^{i\pi}\right\}$. 
An immediate consequence is that the 3-dimensional rotational group can not represent the 
6-dimensional RTR group. This implies that the usual convention in literature (three color scheme as in Fig. 1.b of \cite{NatK}) 
does not represent faithfully the combination of symmetries given by both (\ref{eqHR0}) and (\ref{eqHT0}). 
Fig. \ref{fig1}d shows a complete classification scheme for integer $J$ and $n=3$ with the 6 relevant elements. 

\begin{table*}[!hbtp]
\centering
\caption{\it\footnotesize  
List of all possible $\hat R \mathcal{\hat T}\hat R$-related phases, $e^{i\varsigma}$ (upper  rows)  and 
$\hat R$ eigenvalues, $e^{i2\pi/n}$ (lower rows), for any $n$ and $m$ combination. 
I = integer $m$, HI = half-integer $m$.  
}
\begin{tabular}{|c c|| c| c| c|  c| }
  \hline
Symmetry, &$m$ & $n$ = 2 ($d$=2)  & $n$ = 3 ($d$=6)  & $n$ = 4 ($d$=1) & $n$ = 6 ($d$=6) \\ \hline\hline
$\mathcal{\hat{T}^\pm}$&I & $ \pm 1$  & $\pm 1$, $e^{\pm i\pi/3}$, $e^{\pm 2 i\pi/3}$ & $+1$ & $\pm 1$,$e^{\pm i\pi/3}$, $e^{\pm 2 i\pi/3}$ \\
&HI& $e^{\pm i\pi/2}$ & $e^{\pm i\pi/6}$, $e^{\pm i\pi/2}$, $e^{\pm 5i\pi/6}$& $-1$ &   $e^{\pm i\pi/6}$, $e^{\pm i\pi/2}$, $e^{\pm 5i\pi/6}$\\ 
\hline 
$\hat R$&I & $\pm$1& 1, $e^{\pm i 2\pi/3}$ & $\pm$1,  $e^{\pm i\pi/2}$ & $\pm$1, $e^{\pm i \pi/3}$, $e^{\pm i 2\pi/3}$ \\
&HI & $e^{\pm i\pi/2}$ & -1, $e^{\pm i\pi/3}$ & $e^{\pm i\pi/4}$, $e^{\pm i3\pi/4}$&  $e^{\pm i \pi/6}$, $e^{\pm i \pi/2}$,$e^{\pm i5 \pi/2}$ \\
\hline
\end{tabular}
\label{table1}
\end{table*}
Table  \ref{table1} lists all the possible eigenvalues of the time-reversed pair 
of $\hat R$  and related  TR  phases, for any combination of $m$ and $n$. 

For completeness, we evaluate next the possible protection mechanisms to first order based on the values of table \ref{table1}, 
allowing us to obtaining a general classification scheme. 
In general, the experimental setups designed for  memory storage aim for a magnetic ground state as robust (or long lived) as possible. 
Hence, it appears relevant to study the protection to zero-th order \cite{delgado,romeike}, by examining
 $\langle \psi^{\varsigma(m)}_{m} | \psi^{\varsigma(-m)}_{- m} \rangle$. 
We emphasize that the phases given by $\varsigma(m)$ and $\varsigma(-m)$ are in general different. 
It is immediate to see that the pair is actually the \emph{same} state (i.e., given by the same representative) when
 both $m$ and $2m/n$ are integers. The latter condition indicate that they are the same rotational class, 
whereas the former leads to a real representation of the time-reversal group ($i\varsigma(m)$ is then an integer multiple of $i\pi$). 
In addition, a coefficient relation may easily be obtained,  
using (\ref{eqtr1}) in (\ref{eqpsim}): 
\begin{equation}
\label{coefs}
c_k^m = {\rm e}^{i\pi n k} (c_k^{-m})^*. 
\end{equation} 
In terms of the original components, the crystal Hamiltonian $H_n$ mixes $|J,\pm M\rangle$ and  $|J,\pm(M-n k )\rangle$. 
Together with (\ref{coefs}), and choosing the coefficients $c_k^m$ to be real, 
this implies non-magnetic ($\langle \hat J_z\rangle = 0$) bonding and anti-bonding mixtures, 
commonly termed as quantum spin tunneling splitting \cite{book}. 
This brings two consequences: (i) if $m = \pm J$, then preparation of a magnetic ground state would result into short relaxation times \cite{NatK}
and quantum spin tunneling \cite{rossK} 
(ii) if $m \neq J$, this results in the appearance of a ‘shortcut tunneling’ \cite{NatK,huebner}. 
 
In typical samples, the main mechanism for magnetization reversal are spin-flip events mediated by exchange interaction with 
substrate or tunneling electrons \cite{huebner,lorente,khaje,chudo,rossier, rossier2}, which can be described in its most general form, as 
\begin{equation}
\hat H_e = \alpha \hat{\vec{J}}\cdot\hat{\vec{\sigma}}, 
\label{eqHe}
\end{equation}
where $\sigma_i$ are the usual Pauli matrices denoting the electronic spin and $\alpha$ is an arbitrary constant.
In the weak coupling limit, the degrees of freedom of the electrons and the adatom factorize. 
Using the anti-unitarity of the TR operator and Eqs. (\ref{eqtplus}) and (\ref{eqTpp}), we have  
\begin{eqnarray}
\langle \psi_{m}^{\varsigma(m)} | \hat{\vec{J}}  | \psi_{-m}^{\varsigma(-m)} \rangle &=&  
\langle \mathcal{ \hat T^+}  \psi_{m}^{\varsigma(m)} |
\mathcal{ \hat T^+}  \hat{\vec{J}}
(\mathcal{ \hat T^+})^{-1}  |\mathcal{ \hat T^+}  \psi_{-m}^{\varsigma({-m})} \rangle^* =
\nonumber \\ &=& 
-\langle \mathcal{ \hat T^+}  \psi_{m}^{\varsigma(m)} |\hat{\vec{J}}|\mathcal{ \hat T^+}  \psi_{-m}^{\varsigma({-m})} \rangle^* =
\nonumber \\ &=& 
-{\rm e}^{-2i\varsigma(m)}\langle \psi_{m}^{\varsigma(m)} | \hat{\vec{J}}  | \psi_{-m}^{\varsigma({-m})} \rangle, 
\label{eqForK}
\end{eqnarray}
where we have used  $\langle \mathcal{ \hat T^+}  \psi_{m}^{\varsigma(m)} | = e^{-i\varsigma(m)} \langle \psi_{-m}^{\varsigma(-m)} |$
and $|\mathcal{ \hat T^+}  \psi_{-m}^{\varsigma({-m})} \rangle = e^{i\varsigma(-m)}|\psi_{m}^{\varsigma({m})} \rangle = 
e^{-i\varsigma(m)}|\psi_{m}^{\varsigma({m})} \rangle $. 
We stress the importance of the phase $\varsigma(m)$, since the matrix element may exist only when 
(i) $2i \varsigma(m) = i\pi(2l+1)$, $l\in\mathbb{Z}$, {\it and} 
(ii) $-m+1$ and $m$ (or $-m-1$ and $-m$) belong to the same class, 
or equivalently, when (i) $2m(1+4/n) $ is an \emph{odd integer} and (ii)  $(2m-1)/n$ [or $(2m+1)/n$] is an \emph{integer}. 

For the former, (i),  we would like to stress that a complex phase different to $2i\varsigma(m)= i\pi$ 
forces the matrix element in Eq. (10) to be 0, which we will term as time-reversal protection. 
A geometrical interpretation of this cancellation was already noted by von Delft {\it et al.} \cite{delft}, 
where the topological phase leads to destructive interference between the symmetry-related tunneling paths. 
Although not affecting the overall results in the particular case studied by \cite{NatK, karlewski}, 
we want to note that this phase  was not taken into account by Miyamachi {\it et al.} nor by Karlewski {\it et al.}
yielding a difference between our Eq. (\ref{eqForK}) and Eq. (4) in \cite{NatK} or Eq. (6) in \cite{karlewski}.  
Clearly, condition (i) is never satisfied for their particular  $m$ integer case,
however, their approach would fail {\it e.g.} in the simple case with  $n=2$, and half integer $J$.



The geometric interpretation of this phase can be given in terms of the Euclidean action. 
The general SA Hamiltonian given by $\hat H_s = \hat H + \hat H_e$ (\ref{eqHJ}, \ref{eqHe}) has an easy axis
(say the $z$ axis, $\theta =0$), around which it has $n$-fold symmetry, $\hat H_s(\theta,\phi) = \hat H_s(\theta,\phi+2\pi/n) \ \forall\ \phi$.
Now we consider an initial and final  state  corresponding to the two degenerate classical ground states,
$|\theta =0\rangle$ and $|\theta=\pi\rangle$.
The tunneling amplitude is then given by $\langle\theta=0| e^{i\hat H_s t/\hbar}|\theta=\pi\rangle$.
If  $T(\phi=0)$ is a tunneling path from $|\theta =0\rangle$ to $|\theta=\pi\rangle$
(broken black arrow of \ref{cartoon}), then $T(\phi=-2\pi/n)$ defined by $\hat R_{-\phi} \hat T \hat R_{-\phi}$ 
(blue arrows of  \ref{cartoon}) is also a tunneling path, related by
symmetry to  $T(\phi=0)$. The corresponding tunneling amplitudes differ at most by a topological phase, given by
the Euclidean action $S(T)$ over the symmetry related paths \cite{delft} and often refer to as a Berry phase \cite{haldane},
which reduces to, for our SA:
\[
S[T(\phi=0)] -S[T(\phi=2\pi/n)] = i4\pi \frac{m }{n} 
\]
The total amplitude acquires then a phase given by the sum over all paths. Since the number of paths is given 
by the dimension $d$ of the TRT, we may write
$\Delta\varphi = \sum_{k=0}^{d-1} e^{i4\pi m k/d} $,  which is clearly 0 unless $2m$ is an integer multiple of $d$. 
In this sense, $\varsigma$ can be identified with the discretized  Euclidean action, 
$i\varsigma (m,n) = S[T(\phi=2\pi/n)]$, corresponding to the topological term of the Lagrangian. 

\begin{figure*}[!htb]
\includegraphics[width=0.65\textwidth]{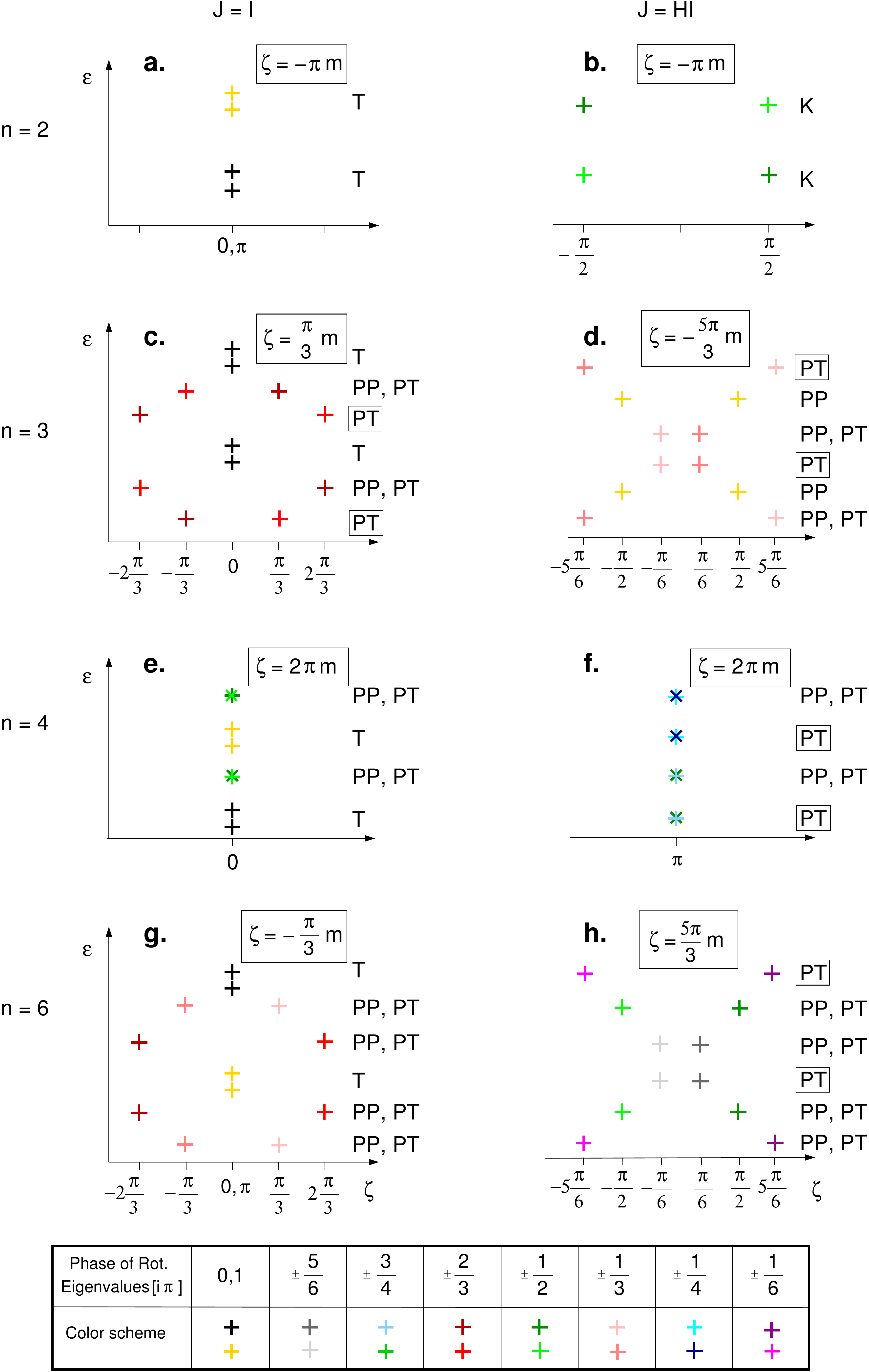}
\caption{
Schematic representation of the lowest energy levels (not to scale) 
as a function of 
$\varsigma(m,n)$ (\ref{eqBerryPhase}), for all possible $m,n$ combinations.  
The different cases are labeled in terms of the possible scenarios: 
T = tunneling, K = Kondo, PP = Protected by point-group, PT = Protected by time reversal. 
Inset: colors are related to the rotational classes. 
}
\label{fig1}
\end{figure*}

Based on the above, we may draw a few conclusions: 
(i) when both $m$ and $2m/n$ are integers,  
we expect non-magnetic energy split states. Otherwise, the pair of states remain degenerate.
(ii) The ground states are protected by both, TR and rotational symmetry for integer $J$ and fractional $2J/n$: 
the latter implies that they belong to different rotational classes. 
One may expect a more robust protection  if $(2J-1)/n$ is also fractional by employing a similar argument. 
(iii) The ground states are protected by TR for a fractional $J$ to zero order, even if $2J/n$ is integer, and  further, 
if $(2J-1)/n$ is fractional, then symmetry protection appears to first order. 
Finally, (iv) the ground states are protected by TR symmetry for fractional $J$ and fractional $2J/n$. 
In this last case, if $(2J-1)/n$ is 0 or integer, then 
the protection is exclusively by TR, and we may say the states are {\it topologically protected}. 
This topologically protected states may occur for $m$ half integer and (a) $n=3,6$ and $m = (1+6k)/2$,  
(b) $n=4$ and $m \neq 5/2, 11/2, \dots (1+4k)/2$, 
(c)$n=3$ and  integer $m=1+3k$ with $k$ integer or 0 (see boxed `PT' on the right side of Fig. \ref{fig1}c, d and h). 

We proceed now to evaluate all the possible symmetry combinations. 
The ordering of the lowest $n$ pairs ($2n$ pairs, for $n=3$) is depicted in Fig. \ref{fig1}, as a function of the phase $\varsigma(m)$ defined in 
(\ref{eqBerryPhase}), 
assuming $|D_z|>E_n$, $D_z<0$ and $J>n$. 
The left panels (a,c,e,g) correspond to integer $J$, whereas the right ones (b,d,f,h) belong to half-integer $J$. 
For each case we consider only the first $l = $max$\{d,n\}$ pairs of states, hence from $m = \pm J$ to $m = \pm(J-l)$.
These describe the first fold, pretty much like the $k$- states in a Brillouin zone. 
In fact, higher energy states will then acquire similar discrete phases.  
{Note} that the ordering of the states could be altered when considering higher order terms, {whereas truncation of the 
fourth-rank ($k$ = 0) terms may introduce errors} \cite{donati,truncation}.  
However, for the sake of simplicity, we restrict our discussion to the Hamiltonian defined in (\ref{eqHJ}).
The labels (PT, PP, K, T) describe the protection of each pair, assuming the pair is at the ground state, 
with T = tunneling, K = Kondo, PP = Protected by point-group, PT = Protected by time reversal (see below).
The colors indicate the rotational phase, {\it e.g.}, for $n=2$,  if spin is half-integer, the rotational phase can only take 
$\pi/2$ or $-\pi/2$, which we label with dark and light green, respectively, whereas it 
take $\pi$ or $2\pi$ in the spin integer case, labeled black or yellow. 

For $n$ = 2 (panels a,b of Fig.  \ref{fig1}), 
noting that 2$m/n = m$ is integer (fractional) whenever $m$ is (semi-)integer, 
we conclude that the two states always belong to the same (different) rotational class.  
For integer $m$, there is only one possibility, which is the non-magnetic combinations of states, commonly termed as the 
tunneling case (T) \cite{rossK}.  
For half-integer $m$, 
the exchange mixing 
is always allowed, since $2i\varsigma(m) = i\pi(2l+1)$, and $(2m\pm 1)/n = l$, $l \in\mathbb{Z}$ . 
This is the so-called Kondo (K) case, where scattering with a single electron leads to 
transitions between the two ground states, as was studied by Ternes {\it et al.} \cite{ternes}. 

For $n$ = 3 (panels c, d of Fig.  \ref{fig1}) and 
$m$ integer, $2m/n$ is integer for $m$ multiple of 3
bearing the split states (black, rotation eigenvalue = 1) already noted in the panel above (T). 
When $2m/3$  is fractional, the RTR eigenvalues are 
$\pm \pi/3$ or $\pm 2\pi/3$, which can have a rotational phase of $\pm 2\pi/3$ (brown and red, respectively). 
If the next level is the split pair, then the pair is protected by both point group and TR symmetry (PP,PT). 
However,  if the next pair contains the same rotational classes, then the single electron scattering mechanism 
is still unavailable, since $2\varsigma(m,n=3)$ is fractional (see Eq. (\ref{eqForK})), 
bearing a cancellation by topologically protected states (labeled PT).  
This is consistent with the protection observed by Miyamachi {\it et al.} \ref{natK}.
For half-integer $m$, $e^{i\varsigma(m,n=3)}$ is 
always complex, implying that Kramers theorem applies for all states. 
Even if the pair belongs to the same rotational class, 
neither direct mixing nor exchange mixing may occur 
(yellow pair of Fig.  \ref{fig1}.d).  
Finally, we find a set of topologically protected states as in the integer case.


For $n$ = 4, $i\varsigma(m) = 2i\pi m$. 
With the restriction $-\pi<\varsigma(m)\le\pi$, all states fall in phase zero ($\pi$)  for integer $m$ (half-integer $m$). 
It is easy to see that if $m$ is an even integer, 
then the pair is in the same rotational class, resulting in the splitting (T), whereas an odd integer $m$ 
results in protected states also under exchange with electrons: the rotational phases of the pair are $\pm \pi/2$. 
This results in the alternating pattern of Fig. \ref{fig1}e, consistent with the protection observed by Donati {\it et al.} \ref{donati2}.   %
In the half-integer case, none of the conditions for exchange with electrons $2i\varsigma(m) = i\pi(2l+1)$, $l\in\mathbb{Z}$ or $(2m-1)/n$ 
integer are satisfied, bearing both PT and PP for all levels. 

Finally, for $n$ = 6  (panels g, h of Fig. \ref{fig1}), six different rotational classes appear, resulting 
in enhancement of the protection of the states. 
First order exchange with electrons is always forbidden. 
For half-integer $J$, the protection is to highest order of all the possible cases. 

Note that the theory presented here applies as well for $J\leq n$, only that in this case not all the states of Fig. \ref{fig1} 
would enter the diagram.
For instance, if we consider $J=3/2$ and $n = 6$, only the four topmost states depicted in Fig. \ref{fig1}h 
would appear. 
A generalization to the case  where the GS is not dominated by $M=\pm J$ is possible with some modifications. 
Although the same phase spectrum $\varsigma(m)$ would be observed, the energetic ordering would change and also 
the evaluation of the matrix elements in (\ref{eqForK}) would need to be re-adapted. 



\section{Conclusions} 
In conclusion, we have derived the quantum numbers that describe the discrete rotations of a nanomagnet
with discrete rotational symmetry. 
In combination with  time reversal symmetry, a non-trivial topological phase is derived,
revealing the intriguing protection mechanisms of the resulting ground states, associated to a rotated 
time reversal operator. 
We have developed a comprehensive classification scheme based solely on symmetry arguments, which reveals 
a particular folding of states. 
Our results are relevant for the proposal of SA-based memory storage devices, as well as for the 
recent studies revealing Kondo effect in SA.  Moreover, our findings should stimulate experimental 
groups to find interferometry experiments that allow to unveil this geometric phase \cite{guido}. 
\section{\bf   Acknowledgments:}  
We thank D. Pfannkuche, C. H\"ubner, A. Chudnovskiy and O. Tcheryshyov for  valuable discussions. 
\section{\bf Additional Information:} 
We acknowledge support of this work by the Deutsche Forschungsgemeinschaft (Germany) within SFB 925 and GrK 1286. 
{\bf Competing financial interests:}
The author declares no competing financial interests.

\end{document}